\documentclass[amssymb,amsmath,aps,floatfix,nofootinbib, preprintnumbers ,12pt]{revtex4}
\usepackage{color}
\usepackage{graphicx}

\def\beq{\begin{equation}}
\def\eeq{\end{equation}}

\def\be{\begin{equation}}
\def\ee{\end{equation}}
\def\bea{\begin{eqnarray}}
\def\eea{\end{eqnarray}}

\newcommand{\gsim}{\lower.7ex\hbox{$\;\stackrel{\textstyle>}{\sim}\;$}}
\newcommand{\lsim}{\lower.7ex\hbox{$\;\stackrel{\textstyle<}{\sim}\;$}}

\begin{document}


\bigskip

\title{Go with the Flow, Average Holographic Universe\footnote{Essay written for the Gravity Research Foundation 2010 Awards for Essays on Gravitation}}

\author{George F. Smoot$^{1,2,3,4,5}$}
\email{gfsmoot@lbl.gov}
\affiliation{$^1$Institute for the Early Universe, Ewha Womans University \& Advanced Academy, Seoul, Korea}
\affiliation{$^2$Institute for the Physics and Mathematics of the Universe,
University of Tokyo, Kashiwa, Chiba 277-8568, Japan}
\affiliation{$^3$Lawrence Berkeley National Lab, 1 Cyclotron Road, Berkeley, CA 94720, USA}
\affiliation{$^4$Physics Department, University of California, Berkeley, CA 94720, USA}
\affiliation{$^5$Chaire Blaise Pascale, Universite Paris Denis Diderot, Paris}

\begin{abstract}
Gravity is a macroscopic manifestation of a microscopic quantum theory of space-time,
just as the theories of elasticity and hydrodynamics are the macroscopic manifestation 
of the underlying quantum theory of atoms.
The connection of gravitation and thermodynamics is long and deep.
The observation that space-time has a temperature for accelerating observers and horizons 
is direct evidence that there are underlying microscopic degrees of freedom.
The equipartition of energy, meaning of temperature, in these modes leads one to anticipate
that there is also an entropy associated.
When this entropy is maximized on a volume of space-time, 
then one retrieves the metric  of space-time (i.e. the equations of gravity, e.g. GR).
Since the metric satisfies the extremum in entropy on the volume,
then the volume integral of the entropy can readily be converted to surface integral,
via Gauss's Theorem.
This surface integral is simply an integral of the macroscopic entropy flow
producing the mean entropy holographic principle.
This approach also has the added value that it naturally dispenses with the cosmological constant/vacuum energy problem in gravity except perhaps for second order quantum effects on the mean surface entropy.
\end{abstract}


\maketitle

\newpage

\section{Introduction}

\bigskip
\noindent In the early 1970's the four laws of black home mechanics, 
which at the time were noted to be extremely parallel to those of thermodynamics,
were originally derived from Einstein equation of General Relativity\cite{Bardeen73}.
Later it became clear, with the semi-clasical derivation of Hawking radiation\cite{Hawking75} and Bekenstein entropy,  that these were the same laws.
How did classical General Relativity know that horizon area is a form of entropy 
and that surface gravity is temperature?

That question was answered in part by a 1995 derivation of General Relativity from
the proportionality of horizon area to entropy, surface gravity to temperature, and the first law of thermodynamics,
in particular $\delta Q = T dS$\cite{Jacobson95}.

This was not the only hint at the time.
The Fulling-Davies-Unruh effect, was first described by Steven Fulling in 1973, Paul Davies in 1975 and Bill Unruh in 1976.\cite{Unruh76} 
It is the prediction that an accelerating observer will observe a thermal bath including black-body radiation where an inertial observer would observe none. 
The ground state for an inertial observer is seen as in thermodynamic equilibrium with a non-zero temperature by the uniformly accelerated observer.
Because of the equivalence principle a uniform acceleration and a uniform surface gravity would be anticipated to agree.
This was clear evidence that acceleration and gravity were related to thermal physics and
equipartition of energy.
The fact that materials could be heated convinced Ludwig Boltzmann in 1876-1890 that there were underlying degrees of freedom and that he must develop a statistical treatment of nature.
Boltzmann's kinetic theory of gases seemed to presuppose the reality of atoms and molecules, 
but at the time almost all German philosophers and many scientists like Ernst Mach and the physical chemist Wilhelm Ostwald opposed their existence.

Things moved another step forward into a cosmological context 
with the derivation of the Friedmann equations from the
first law of thermodynamics\cite{CaiKim05} on the apparent cosmic event horizon
and the assumption that the entropy is proportional to the area and temperature to the surface gravity.

In parallel with this increasing solidification of the links between gravity and thermodynamics,
there has been increased attention to entropy limits beginning in 1981 with Berkenstein's bound\cite{Bekenstein81}
setting a universal maximum to the entropy enclosed within a volume.
This lead eventually to the holographic principle\cite{'tHooft93, Susskind95},
which was made covariant and precisely defined in 2002\cite{Bousso02} through the use of the volume 
being defined by the two convergent light sheets from the boundary surface area.
There is strong evidence that the area of any surface limits the information content of adjacent space-time regions, at $1.4 \times 10^{69}$ bits per square meter.  
A universal relation between gravitation, geometry and information was thus uncovered. 
It has yet to be explained clearly. 
The holographic principle's origin must lie in the number of fundamental degrees of freedom involved in a unified description of space-time and matter or in a regulation of allowed configuration of states. 

The holographic principle states that the description of a volume of space can be thought of as encoded on a boundary to the region, preferably a light-like boundary like a gravitational horizon. 
First proposed by Gerard 't Hooft\cite{'tHooft93}, it was given a more precise string-theory interpretation by Leonard Susskind\cite{Susskind95}.
In a larger and more speculative sense, the theory suggests that the entire universe can be seen as a two-dimensional information structure ``painted" on the cosmological horizon, such that the three dimensions we observe are only an effective description at macroscopic scales and at low energies. 
Cosmological holography has not been made mathematically precise, partly because the cosmological horizon has a finite area and grows with time.
The holographic principle was inspired by black hole thermodynamics, which implies that the maximal entropy in any region scales with the radius squared, and not cubed as might be expected. In the case of a black hole, the insight was that the description of all the objects which have fallen in can be entirely contained in surface fluctuations of the event horizon. The holographic principle resolves the black hole information paradox within the framework of string theory.

\noindent Here we turn this logical development  on its head. 
That is start with a principle and derive a theory that describes what we observe.

\subsection{The Conjecture}
I conjecture that: {\bf All of the possible histories of the universe, past and future, are encoded on the apparent horizon of the universe.}

We take the point of view that classical and even semi-classical space-time is the coarse-grained limit of microscopic structure of this averaged information.
\noindent

\subsection{What is enough entropy to handle all the past history of the Universe?}

Seth Lloyd\cite{Lloyd05} generated a formula the number of operations $N$
\footnote{ A naive estimate is much larger.  If the universe starts with a relative small number of bits, say $10^{10}$
and we allow one calculation step per Planck time $t_{Pl} = 5.3924(27) \times 10^{-44}$ s  per site which would be a Planck volume.  
We would find the number of calculations $N = \frac{4 \pi r^3 t}{3 \ell_{Pl}^3 t_{Pl} }$
which is roughly $(r/\ell_{Pl})^2$ times larger.
We take an age of the universe as
$T_{universe} = 13.7 \times 10^9 ~years~ = 4.32 \times 10^{17} ~sec~ = 0.80 \times 10^{62}$ Planck times.
and the horizon as roughly $r = 10^{10} light-years =  10^{26} m = 6 \times 10^{70} \ell_{Pl }$.
Why is it so much larger? Because Lloyd uses the black hole production limit to go from $r^3$ to $r$.
We can be a bit more careful and ask how many degrees of freedom could there be in a volume $V$?
We would estimate that there are $V/V_{Pl} \equiv V/\ell_{Pl}^3$ possible sites. 
The minimum number of independent quantum states is $n = 2$ (e.g. a simple spin up and spin down system).
The total number $N$ of independent quantum states in a specified volume $V$ is then $N \sim n^{V/V_{Pl}}$. 
The entropy should be $S/k_B = ln(N) = \frac{V}{V_{Pl}} ln(n)$.
In a time $t$  there are $t/t_{Pl}$ possible state changes. 
Giving us a computation total possible of  $C = \frac{t V}{t_{Pl} V{Pl}}$.
This means we have a reduction of order $l^2/\ell_{Pl}^2$ for calculations and $\l/\ell_{Pl}$ for degrees of freedom. }
 or events that can have taken place in volume with radius $r$ over a time $t$ is 
\begin{equation}
N =  \frac{r t}{\ell_{Pl} t_{Pl}}  = \frac{10^{26} ~m \times 4.32 \times 10^{17} ~sec}{1.616252(81) \times 10^{-35}~ m \times   5.3924(27) \times 10^{-44} s} = 0.5 \times 10^{122}
\end{equation}

We can compare this to the number of bits set by the entropy of a horizon
\begin{equation}
S_H = \frac{k_B c^3}{G \hbar} \frac{A}{4} = \frac{k_B c^3} {G \hbar} \pi R_H^2 = \frac{k_B c^3} {G \hbar} \pi \left( \frac{c}{H} \right)^2 \sim (2.6 \pm 0.3) \times 10^{122} k_B
\end{equation}
Note that these two agree to an amazing factor of order unity.
The relationship one expects is that $S = N ln(2) k_B$ where $ln(2) = 0.693$, so that the factor is
$F = 2.6  /(0.693 \times 0.5) = 7.5$.

If the conjecture is true then we can estimate how much time we have left.
It has to be of the order of a few times what has elapsed.
This is not surprising, if we live in an accelerating rate of expansion universe.
The conformal time that will elapse is roughly
\begin{equation}
\Delta \eta = \int_{t = now}^\infty \frac{dt}{a(t)} \sim \int_{t = now}^\infty \frac{dt}{e^{Ht}} = - \frac{1}{H} e^{-H t} |_{t=now}^\infty = \frac{1}{H} \simeq  15 \times 10^{9} years
\end{equation}
where we take $a(now) = 1$.
This is the same as the allowed past conformal time as one would expect from the light sheet approach.
By having acceleration at the beginning and the end of the universe, 
one can have apparently infinite local physical time.

Note that for an accelerating universe,  the total number of calculations for the entire history and future of our part of the universe (inside our apparent horizon) is only of order a factor of two beyond what has already occurred and can readily fit on the holographic surface, if the calculational time is conformal time.
We might turn that argument around and note that for the total information to fit upon the apparent horizon, 
we need for the universe to accelerate. 
One problem with this argument is that when the universe was much younger, 
the entire horizon was much smaller.
How and why all the information about the observable universe could fit on a smaller horizon?
The answer is that the horizon just contains the information about the history and future of its contents.
When the universe was much smaller, it had not so much contents - though it is still extremely empty now.
Because of that it was possible on a much smaller horizon to keep track of the history and future of that epoch's contents. 

When the universe goes through a decelerating phase more information can enter the universe, which also results in a larger horizon and more entropy/information just as more calculations are necessary.

\subsection{Analogy}
One can think of the universe as a huge holodeck\footnote{ A holodeck is a simulated reality facility located on starships and starbases in the fictional Star Trek universe. It uses very realistic but some times shimmering holograms to make it 3-D and there is even a holographic doctor. The Doctor, an Emergency Medical Hologram Mark I (or EMH for short) is a fictional character from the television series Star Trek: Voyager, played by actor Robert Picardo. 
The EMH is a holographic computer program designed to treat patients during emergency situations, or when the regular medical staff is unavailable or incapacitated. Programmed with all current Starfleet medical knowledge, the Doctor is equipped with the knowledge and mannerisms of historic Federation doctors, as well as the physical appearance of his programmer, Dr. Lewis Zimmerman.} virtual reality stage for a many player game.
We think of the  program and the information for the game or the simulation as the information necessary to define the virtual reality universe.
In the case of games and the holodeck there is undoubtedly some basic world model information (e.g. images and models of things) and a set of rules. 
There are also input sensors that get limited input from the players and have decision rules as to which path to take next and run numerical simulations to provide the next scene.
All of that is stored in what is probably a surface memory; holographic in the sense 2-D storage of information then recreates the history of a 3-D world.
There is little doubt that fairly realistic virtual reality games and movies exist and that they are improving.

\noindent
This is a bit more interactive than I am anticipating for the universe's case.

\noindent
Instead consider a more limited situation: a checkers or chess match.
Here simply calculate every possible position of every possible game and store that
so that when a player wants to play the virtual reality opponent,
the VR opponent simply follows down the memory of all possible games looking ahead with
some weighting function, based upon outcomes, to decide what is the best good move.

\subsection{Punchline of Analogy}
Why do I make this analogy,
Since 1948 the Feymann path integral formulation (started by Dirac, in 1938)
basically calls for the evaluation over all possible histories and then
the path integral simply calculates the probability amplitude for any given process by
summing over all possible paths weighted equally. 
After a long enough time, interference effects guarantee that only the contributions from the stationary points of the action give histories with appreciable probabilities.

The path integral also relates quantum and stochastic processes, and this provided the basis for the grand synthesis of the 1970s which unified quantum field theory with the statistical field theory of a fluctuating field near a second-order phase transition. 
What is needed here is a synthesis of quantum and thermal fluctuations and perhaps space-time emergence  is a second-order phase transition.

Thus having all the information of all possible histories and futures of the universe
is very convenient and useful for determining the actual events of the universe.
In this case the multiverse is us. 

\section{Equipartion and its Role}

One method of doing the path integral is to visit every point in space-time (or in this case data storage location on the surface) with equal weighting.
Then the sum of all the amplitudes from all the visits visits is simply the integral in units of Planck areas.
In the case of the path integral we are simply summing a phase from each possible path (space-time events) and looking for the stationary points. 
Those would be the points that were stationary under the variation of all possible histories (paths through space time).
This also has a direct relationship to the entropy of the system.
What do we mean by equipartition? 
We mean that the system visits every possible state with equal probability.

The validity of the continuum approximation may be verified by a theoretical analysis, 
in which either some clear periodicity is identified or statistical homogeneity and ergodicity of the microstructure exists. 
We can note that we are averaging over scales that give very many samples.
In one meter we have $10^{35}$ samples and we are averaging that either by area (squared) or by volume (cubed).
The smallest distances we have probed with accelerators corresponds to roughly $10^{-18}$ m which is 
averaging over $10^{17}$ either cubed or squared depending upon whether one
averages over volumes or areas.

\section{Coarse-Graining of the Universe to Gravity and Entropy}
We shall follow the lead of Boltzmann and assume that there are microscopic degrees of freedom
on a very small scale. 
These degrees will take very high energy to excite from one state to another,
much like excitation of an atom's electron levels (except the outer most bandgap electrons) and nuclear and subnuclear energy states.
There are then available some collective modes - much like a condensed matter lattice having phonons etc.
In this case these correspond to matter and energy (particles and fields) and the 
deformation of the metric.
The entropy of any given relatively small volume of space is set by the sum of the entropy of the microstructure plus the entropy of the collective modes, which are what we in the past have regarded as the primary entropy content of space time.

The Gibbs Equation for matter and heat flow in the entropy representation is
\begin{equation}
dS = \frac{1}{T} dU + \frac{P}{T} dV - \sum_j \frac{\mu_j}{T} d N_j ~~~~~~ or ~~~~~ d s = \frac{1}{T} du - \sum_j \frac{\mu_j}{T} d c_j
\end{equation}
where the second holds for a fixed volume and  $s = S/N$, $u = U / N$, and $c_j = N_j /N$; $N_j$ are the various chemical species and $\mu_j$ are their respective chemical potentials.

We need to define the continuity equations for energy and entropy in a region small enough to be in local equilibrium.
Define the energy flow vector ${\bf J_U}$  and the entropy flow vector $ {\bf J_S}$.
The continuity equation  for energy and entropy are 
\begin{equation}
\frac{du }{dt} + {\bf \nabla \cdot J_u} = 0          ~~~~    \frac{d s}{d t} + {\bf \nabla \cdot  J_S} = \Phi 
\end{equation}
where $\Phi = \sum_j \frac{\mu_j}{T} \bf{ \nabla \cdot J_{N_j}}$ is the rate of entropy density production in the volume.

We find the global relations by integrating the local relations over the volume and using the Divergence Theorem:
$\int_V {\bf \nabla \cdot J_S} = \int_\Sigma  {\bf J_S \cdot d A}$ to find
\begin{equation}
\frac{d}{dt} \int_V s ~dV +\int_\Sigma {\bf J_S \cdot d A} = \int \Phi dV = \Theta
\end{equation}
The right hand side of this equation is the rate of entropy production $\Theta$ in the full volume.
If there is no internal entropy production, or the entropy is maximized (at extremum), 
then there is a direct relation between the volume entropy and the 
surface integral of the entropy flow.
We are used to this in a conserved charge situation; e.g.
$\iint_\Sigma {\bf E \cdot d A} = \sum_i q_i$ where $q_i$ are the charges enclosed by closed surface $\Sigma$. 
\footnote{Gravity has the corresponding requirement on the Lagrangian $\sqrt{-g}L_{surface} = -\partial_a \left( g^{ij}\frac{\partial \sqrt{-g} L_{Bulk}}{\partial \partial_a g_{ij}}\right) $  \cite{Padmanabhan10} and one can get the equations of gravity from the bulk ignoring the surface terms and varying the metric or from the surface terms ignoring the bulk and not varying the metric but varying the surface and finding the gravity equations as a result.}

To make this classical chemical entropy formulation covariant one must use tensor formalism and find the general expression for the entropy density and the entropy flux.
Achieving this and maximizing (calculus of variations extremal) the entropy makes it clear where generic equations of gravity arise and why one must us the time-sheet definition for the holographic bound.

Jacobson\cite{Jacobson95} extended by Padmanabhan\cite{Padmanabhan10} have shown that the simplest form for the entropy density  and the entropy flow is simply leads to the Einstein Equations of GR $\left( R_{\mu \nu} - \frac{1}{2} R g_{\mu \nu} + \Lambda g_{\mu \nu} - 8 \pi G T_{\mu \nu} \right) n^\mu n^\nu$ where $n^\mu$ is an arbitrary null vector. 
When at an entropy extremum  is taken for all possible null vectors this must be zero and the surface entropy term is what remains.
Thus GR emerges as the proper deformation of space-time to maximize the entropy of space-time given its matter-energy content (stress-energy tensor). 
One can think of various sensible extensions to the definition of the entropy density tensor and they lead to additional forms of gravity such as Lanczos-Lovelock with a similar surface term.
In this form, gravity is actually another branch of continuum mechanics like elasticity and fluid flow. 
Gravity's equations are an equation of state for space-time.

A key issue is that the surface entropy term is well-defined upon  the light sheet enclosing the volume as one might expect for causality.
As such it averages over all the quantum fluctuations in the volume along the light cone, whose means are nominally zero on average except for the focusing by the matter-energy content and statistical fluctuations.
One gets the dimensional reduction by averaging the volume (e.g. past light cones) to each site on the surface.
By choosing the light sheet, which is cut off at caustics, one automatically gets a map of all the contents of the volume and thus up holds the holographic nature.
Were the light rays to cross, so would the entropy flux and that would interfere with the divergence theorem.
This means that the deformation of space-time by matter-energy content in a way to make entropy an extremum 
also mean that in the volume defined by the light sheet, each and every portion of the content is visible
to the surface recording its information holographically.
This is precisely the constraint that one needs for light holography to work - 
the interference pattern recorded on the surface must have free light travel to reconstruct the volume it represents.
For actual holograms this happens automatically as they are generally created by illuminating
the contents of a limited spatial region with clear line of sight light cones to the recording surface.

\section{Why is Temperature proportional to the Curvature?}

When one looks at it, it seems that having the temperature proportional to the curvature of space time is a strange thing.
Why too does the number of states available vary inversely to the curvature?
We have entropy proportional to the inverse of the curvature ($S \propto R^2$) and temperature proportional to the curvature ($T \propto 1 / R^2$) so that the product is a constant for a fixed enclosed energy.

There are many ways to look at this issue.
A simple and historic approach is to consider quantum mechanical tunneling through the horizon.
Several papers have dealt with this issue for Rindler, Schwartzschild, FRW and more exotic horizons
(see \cite{Zhu10} for example and references especially relevant to FRW).
The interesting result is that the tunneling probability is familiar
\begin{equation}
\Gamma \propto e^{Im( \oint p dr ) } \propto e^{- \Delta S_{int}} \propto e^{- \Delta E / k_B T_H}
\end{equation}
where the first is the action and the second is the change in Entropy $\Delta S_{int}$ internal to the horizon in units of $k_B$ and $T_H$ is the effective horizon temperature which is easy to show is given by
\begin{equation}
T_H = \frac{ \hbar }{2 \pi k_B} \frac{c}{r_A} = \frac{ \hbar }{2 \pi k_B} \left( H^2 + \frac{k c^2}{a^2} \right)^{1/2}
\end{equation}

\section{Toy Model}
 Flat space has the configuration that there is a periodic in space and time potential minima all at the same default value which we label zero.
Curve space and that raises the minima  which in turn reduces the number of available states at temperature $k_B T$.
In this model a 2-D spherical surface of radius $r$ would have a number of accessible sites that are given by the area $A = 4 \pi r^2$ and the mean rise in energy level of those states will be proportional to $1/r$.
Thus for a given total energy the number of bits on the surfaces is proportional to  $r^2$ and the temperature is proportional to $1/r^2$ for a fixed enclosed energy or $1/r$ for the fluid-filled FRW universe.

To get this behavior the potential for space-time must look something like
\begin{equation}
V(r) = \frac{1}{2} \alpha m_{Pl} c^2 \frac{r^2}{\ell_{Pl}^2}  + \frac{1}{2} \beta m_{Pl} c^2 \frac{\ell_{Pl}^2}{r^2}
\end{equation}
where the first term comes from the elasticity of space  (cf. GR\cite{Padmanabhan04}) and the second term could come from the Pauli exclusion principle forces for the non-commutative geometry (see Appendix) or for discrete quantum sites.
Differentiating with respect to $r$ we have
\begin{equation}
\frac{d V(r)}{d r}  =  \alpha m_{Pl} c^2 \frac{r}{\ell_{Pl}^2}  -   \beta m_{Pl} c^2 \frac{\ell_{Pl}^2}{r^3}
\end{equation}
This tells us the space-time forces
and setting this to zero gives us the equilibrium point for each cell radius
\begin{equation}
r = \ell_{Pl} \left( \frac{ \beta}{\alpha}  \right)^{1/4}  =   \ell_{Pl} 
\end{equation}
where the last happens if we set $ \beta = \alpha$ and this gives us an area that is roughly $4 A_{Pl}$ for our cells.  
We can determine the actual coefficient $\alpha$ by resorting to General Relativity or Newtonian gravity.
Just like the underlying theory of atomic lattice potentials should give the coefficient of elasticity of deformable solids.
In this picture gravitation is simply like Hooke's law and the theory of deformable elastic solids.

\subsection{More Model Explanation}
In this model to get the mean number of sites to be $A/ A_{Pl}$ we have to have the energy of each state of curved space to be proportional to the curvature ($E_0  \propto 1/r^2$) 
and the mean temperature $k_B T \simeq E_0$.

If for example we take there to be three spatial dimensions then
the partition function gives us for a harmonic oscillator potential
\begin{eqnarray}
Z(T) &=& \sum_{space ~sites} \sum_{energy ~levels ~ n =0}^{n \rightarrow \infty} \frac{ (n+1)(n+2)}{2} e^{-(n+1/2)E_0/k_BT} \cr
&=& \frac{A}{4A_{Pl}} \sum_{energy ~levels ~n =0}^{n \rightarrow \infty} \frac{ (n+1)(n+2)}{2} e^{-(n+1/2)E_0/k_BT}
\end{eqnarray}
and we need this to be 1 for the statistical average to get the right occupancy number.
This immediately gives us $k_BT \simeq E_0$.

While the energy is then
 \begin{eqnarray}
E &=& \sum_{space ~sites} \sum_{energy ~ levels ~n =0}^{n \rightarrow \infty} \frac{ (n+1)(n+2)(n + 1/2)}{2} E_0e^{-(n+1/2)E_0/k_BT} \cr 
&=& \frac{A}{4A_{Pl}} \sum_{energy~ levels ~ n =0}^{n \rightarrow \infty} \frac{ (n+1)(n+2)(n+1/2)}{2} E_0e^{-(n+1/2)E_0/k_BT} \cr
&=& E_0  \frac{A}{4A_{Pl}} \sum_{energy ~levels ~n =0}^{n \rightarrow \infty} \frac{ (n+1)(n+2)(n+1/2)}{2} e^{-(n+1/2)E_0/k_BT} 
\end{eqnarray}
We can see from the last that we have to have the energy level $E_0 \propto 1/A = 1 / (4 \pi r^2)$ to conserve the enclosed energy $E$ for a fixed set of matter and energy enclosed.
Whereas in a cosmological context (and a black hole context) one would expect that the total energy enclosed would increase proportional to  $r \propto M$, and that would give a rise of $E_0 \propto 1/r$.

\section{Discussion}\label{sec:dis}
An actual hologram of an object is made by interfering light from a reference beam and the light scattered by an object.
To make this work effectively, 
a light source goes through a beam splitter to create the coherent illuminating beam and the reference beam in order to maximize the possible interference effects.

How would this happen for the Universe?
Well the entropy flux takes all possible paths on the light sheet including the past and the future light sheet.
This means for the central objects there is information from the most distant past and a roughly equal future epoch which match (modulo the flux through the horizon) at the apparent horizon.
If one moves to a new location away from the central object, then it will have an apparent event horizon centered on it
and have the same information available to it.

We live in the best of all possible worlds because it is all possible worlds and the only world.
That is how it is written in the big book, the apparent horizon.
\bigskip

\begin{center}

\section*{Acknowledgements}

\noindent  
This work supported in part by by the U.S. Department of Energy under Contract No. DE-AC02-05CH11231, by WCU program of NRF/MEST (R32-2009-000-10130-0),
and by CNRS Chaire Blaise Pascal.

\end{center}

\newpage
\appendix{Appendix: Non-Commutative Geometry for the Atoms of Space-Time}

In order to have a well-defined natural segmentation into cells of size given by the Planck length, area and volume, we introduce a non-commutative geometry.
\begin{equation}
\left[ x^\mu, x_\nu \right]  = i \ell_{Pl}^2 \left( 1 - \delta^{\mu}_{\nu}  \right)
\end{equation}
is the simplest possible commutation. It could be more complicated and be an operator with the same norm.
We could have simply defined a discrete set of points and quantum mechanics and states for each of these points and then coarse grain  average them to continuous medium. 

We also have to introduce a new exclusion principle a la Pauli.\footnote{
In quantum mechanics, the spin-statistics theorem relates the spin of a particle to the particle statistics it obeys. The spin of a particle is its intrinsic angular momentum (that is, the contribution to the total angular momentum which is not due to the motion of the particle).
 All particles have either integer spin or half-integer spin (in units of the reduced Planck constant $\hbar$).
The theorem states that:
the wave function of a system of identical integer-spin particles has the same value when the positions of any two particles are swapped. 
Particles with wavefunctions symmetric under exchange are called bosons;
the wave function of a system of identical half-integer spin particles changes sign when two particles are swapped. 
Particles with wavefunctions anti-symmetric under exchange are called fermions.
Spin statistics theorem implies that half-integer spin particles are subject to the Pauli exclusion principle, while integer-spin particles are not. 
Only one Fermion can occupy a given quantum state at any time, 
while the number of bosons that can occupy a quantum state is not restricted.}

This, of course, immediately creates a quantization in space-time cells
in the same way that the non-commutative geometry of Poisson phase space does
i.e. $ \left[ x^\mu, p_\nu \right] = i \hbar$ does.
Note that if $\left[ \hat A , \hat B \right] = i \hat C$, 
then the uncertainties in  quantities $\hat A$ and $\hat B$, defined as
$\Delta A^2 = \left< A^2 \right> - \left< A \right>^2$, obey the relation  
\begin{equation}
\Delta A \Delta B \geq \frac{1}{2}  \left< C \right>  ~~~~ so ~~~~~ \Delta x^\mu \Delta x_\nu \geq \frac{1}{2}\ell_{Pl}^2 \left( 1 - \delta^{\mu}_{\nu}  \right) 
\end{equation}

It does leave us with a modified metric which is important for several reasons not the least important is that we need to define our information and computation covariantly a la Bousso.
Nominally we have a Riemannian metric so that proper distances are defined according to

\begin{equation}
 (c d \tau)^2 = -ds^2 =  g_{\mu \nu} dx^\mu dx^\nu = g_\mu^\nu dx^\mu dx_\nu = g^\mu_\nu dx_\mu dx^\nu = g^{\mu \nu} dx_\mu dx_\nu
 \end{equation}
 where the Einstein summation convention applies for repeated indices and greek indices go from 0 to 4.
 One immediately sees that the order now matters and $g_{\mu \nu}$ which is normally a fully symmetric matrix is now antisymmetric, e.g. in simplest 2D case $c^2 d \hat \tau^2 = c^2 d \hat t^2 - d \hat x^2 + c d\hat t d\hat x - d \hat x c d \hat t$ and a raising and lowering operator is more complicated.
 We can fix this some as
\begin{equation}
 (c d \tau)^2 = -ds^2 = dx^\mu  g_{\mu \nu} dx^\nu = dx^\mu g_\mu^\nu  dx_\nu = dx_\mu g^\mu_\nu  dx^\nu = dx_\mu g^{\mu \nu}  dx_\nu
 \end{equation}


\begin{thebibliography}{999}
\bibitem{Bardeen73}
J.M. Bardeen, B. Carter, and S.W. Hawking, {\it Comm. Math. Phys.} {\bf 31} 161 (1973)
\bibitem{Hawking75}
S. W. Hawking, {\it Comm. Math. Phys.} {\bf 43} 199 (1975)
\bibitem{Jacobson95}
T. Jacobsen ``Thermodynamics of space-time: The Einstein Equation of State'' Phys. Rev. Lett. {\bf} 1260 (1995)
\bibitem{Unruh76}
S.A. Fulling, Phys. Rev. D7, 2850 (1973); P.C.W. Davies, J. Phys. A8, 609 (1975); W.G. Unruh, Phys. Rev. D14, 870 (1976).
\bibitem{CaiKim05}
R.-G. Cai, S.P. Kim, J. High Energy Phys. 0502 (2005) 050.
\bibitem{Bekenstein81}
J. D. Bekenstein, ``A Universal Upper Bound On the Entropy to Energy Ratio for Bounded Systems'', Phys. Rev. D {\bf 23} (1981) 287.
\bibitem{'tHooft93}
G. 't Hooft, ``Dimensional reduction in quantum gravity''. arXiv:gr-qc/9310026
\bibitem{Susskind95}
L. Susskind, ``The World As A Hologram,'' J. Math. Phys. {\bf 36}, 6377 (1995)
\bibitem{Bousso02}
R. Bousso Rev. Mod. Phys. 74, 825 (2002)
\bibitem{Lloyd05}
Seth Lloyd, ``Quantum limits to the measurement of spacetime geometry'',  arVXv:quant-ph/0505064v1
\bibitem{Padmanabhan04}
Padmanabhan, T. ``Gravity as Elasticity of SpaceTime", International Journal of Modern Physics D, Volume 13, Issue 10, pp. 2293-2298 (2004)
\bibitem{Padmanabhan10}
Padmanabhan, T. ``Thermodynamical Aspects of Gravity: New Insights'', arXiv:0911.5004v2
\bibitem{Zhu10}
T. Zhu, J. Ren, D. Singleton, ``Hawking-like radiation as tunneling from the apparent horizon in a FRW Universe'',
arXiv:0902.254v2

\end{thebibliography}
\end{document}